\documentclass[prl,twocolumn]{revtex4}\usepackage{psfrag}
\usepackage{epsfig}
\usepackage{amsmath}
\usepackage{amssymb}

\usepackage[colorlinks=true,linkcolor=blue,citecolor=blue]{hyperref}

\newcommand{\bra}[1]{\ensuremath{\langle#1|}}
\newcommand{\ket}[1]{\ensuremath{|#1\rangle}}

\newcommand{\comm}[2]{\left [ #1, #2 \right]}
\newcommand{\be}{\begin{equation}}
\newcommand{\ee}{\end{equation}}
\newcommand{\avg}[1]{\ensuremath{\langle #1 \rangle}}

\newcommand{\tr}{\textrm{tr}}
\newcommand{\im}{\text{i}}

\newcommand{\ie}{{\it i.e.}}

\newcommand{\eg}{{\it e.g. }}

\newcommand{\etal}{{\it et al.}}

\newcommand{\adop}{\hat a^{\dagger}}
\newcommand{\aop}{\hat a}

\newcommand{\xop}{\hat x}
\newcommand{\pop}{\hat p}

\begin{document}

\title{Large Quantum Superpositions and Interference of Massive Nanometer-Sized Objects}

\author{O. Romero-Isart$^{1}$}
\author{A. C. Pflanzer$^{1}$}
\author{F. Blaser$^{2}$}
\author{R. Kaltenbaek$^{2}$}
\author{N. Kiesel$^{2}$}
\author{M. Aspelmeyer$^2$}
\author{J. I. Cirac$^1$}
\affiliation{$^1$Max-Planck-Institut f\"ur Quantenoptik,
Hans-Kopfermann-Strasse 1,
D-85748, Garching, Germany.}
\affiliation{$^2$Vienna Center for Quantum Science and Technology, Faculty of Physics, University of Vienna, Boltzmanngasse 5, A-1090 Vienna, Austria. }

\begin{abstract}
We propose a method to prepare and verify spatial quantum superpositions of a nanometer-sized object separated by distances of the order of its size. This method provides unprecedented bounds for objective collapse models of the wave function by merging techniques and insights from cavity quantum optomechanics and matter wave interferometry. An analysis and  simulation of the experiment is performed taking into account standard sources of decoherence. We provide an operational parameter regime using present day and planned technology.

\end{abstract}

\maketitle

\begin{figure}[t]
\includegraphics[width=\linewidth]{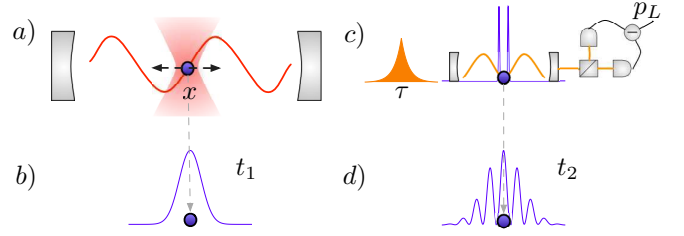}
\caption{(Color online) Schematic representation of the proposal. $(a)$ The optically trapped object is laser cooled using a high-finesse optical cavity. $(b)$ The trap is switched off and the wave function expands during some time $t_1$. $(c)$ The object enters into a second small cavity where a pulsed (of time $\tau$) interaction is performed using the quadratic optomechanical coupling. The homodyne measurement of the output phase measures $\hat x^2$ and prepares a quantum superposition state conditional of the outcome result $p_L$. $(d)$ The particle falls during a time $t_2$ until its center of mass position is measured, which after repetition unveils an interference pattern for each $p_L$.}
\label{Fig:Fig1}
\end{figure}

Quantum superpositions of a massive object at two spatial locations are allowed by quantum mechanics. 
This puzzling prediction has been observed in seminal matter wave interferometry experiments with electrons, neutrons, atoms and dimers, van der Waals clusters, and even complex molecules (\eg $\text{C}_\text{70}$, $\text{C}_\text{60} \text{F}_\text{48}$)~\cite{MatterWave}. Preparing quantum superpositions of even larger objects is considered to be extremely challenging due to the decoherence caused by interaction with the environment~\cite{BooksDec}. 
However, succeeding in this task would allow completely new tests of quantum mechanics: this includes experiments in a hitherto unachieved parameter regime where collapse theories predict quantum mechanics to fail~\cite{Adler2009, Bassi2003}, or even more general tests of quantum theory against full classes of macrorealistic theories~\cite{Leggett1985}.
 Moreover, these states would be so fragile to environmental interactions that one could exploit this ultra-high sensitivity to design a new generation of sensors. Pushing large objects to the quantum regime is also the aim of cavity quantum optomechanics~\cite{ReviewOM}. Similarly to laser cooling of atoms, the radiation pressure of light is exploited to cool and coherently manipulate the mechanical motion of some degree of freedom (\eg the center of mass) of a massive object and even to create quantum superpositions of harmonic vibrational states~\cite{Marshall2003, Romero-Isart2010b}.

In this Letter, we present a method to prepare spatial quantum superpositions of massive objects (with masses of $\sim10^7$ amu) based on cavity quantum optomechanics and show how it can be used to test wavefunction collapse models. This builds upon the recent proposal of using an optically levitating nano-dielectric as a cavity quantum optomechanical system~\cite{Romero-Isart2010b, Chang2010, Romero-Isart2011, Barker2010a}. The main idea is to trap a dielectric sphere in the standing wave field of an optical cavity. The mechanical motion of the sphere's center of mass along the cavity axis is predicted to be a high-quality mechanical oscillator due to the absence of clamping losses. This facilitates laser cooling to its motional ground state (see also experiments on feedback cooling of an optically levitated microsphere~\cite{Li2011}). In addition, a cooled levitating object offers the possibility to be released by switching off the trap~ \cite{Romero-Isart2011}, creating in this way a scenario similar to matter wave interferometry experiments. Here, we will use precisely this feature both to  coherently expand the wave function over a large spatial region and to enhance the non-linear coupling that is required to prepare large quantum superpositions. 

More specifically,  the linear and quadratic coupling in cavity optomechanics after displacing the cavity field (see \eg Sec.~V.A.1 and App.~B.2 in~\cite{ Romero-Isart2011}) is given by
\be
\hat H_\text{OM}= -\hbar g (\aop + \adop) \tilde x + \hbar g_q (\aop + \adop) \tilde{x}^2, 
\ee
where $\aop$($\adop$) is the annihilation (creation) operator of a cavity photon, $\tilde x=\xop/x_0$ is the dimensionless position operator of the mechanical resonator, with $x_0=[\hbar/(2 m \omega_t)]^{1/2}$ its zero point motion, $m$ the mass, and $\omega_t$ the mechanical frequency. The photon-enhanced linear optomechanical coupling is given by $g$ and the typical quadratic coupling by $g_q = k_c x_0 g$, where $k_c=2 \pi /\lambda_c$ is the wave number of the cavity mode. When the equilibrium position of the mechanical oscillator is at the node (anti node) of the standing wave,  $g\neq 0$ and $g_q=0$ ($g= 0$ and $g_q\neq 0$). A fundamental figure of merit of the cavity-mechanical system is the cooperativity parameter defined as
$\mathcal{C}_l = g^2/(\kappa \Gamma)$ for the linear coupling, and  $ \mathcal{C}_q = g_q^2/(\kappa \Gamma)= \mathcal{C}_l \times (k_c x_0)^2$ for the quadratic one. Here, $\kappa$ is the decay rate of the cavity field and $\Gamma$ the decoherence rate of the mechanical motion. 
Ground-state cooling requires $\mathcal{C}_l \gtrsim 1$, whereas non-linear effects,  such as energy quantization detection~\cite{Thompson2008} or preparation of non-Gaussian states without using hybrid systems or single photon resources, require $\mathcal{C}_q \gtrsim1$. The latter is a very demanding condition due to the strong reduction given by $(k_c x_0)^2 \ll 1$. In this Letter we propose to achieve this challenging regime by expanding the wave function to a given size $\avg{\xop^2} \sim \sigma^2 \gg x_0$, such that
\be \label{eq:cooperativity}
\bar{ \mathcal{C}}_q = \frac{\bar{g}_q^2}{\kappa \bar \Gamma} = \mathcal{C}_l \times (k_c \sigma)^2,
\ee
where $\bar g_q$ and $\bar \Gamma$ are defined below. Thus, for sufficiently large $\sigma$ and $\mathcal{C}_l$, the non-linear regime  $\bar{ \mathcal{C}}_q \gtrsim 1$ can be attained. We remark that this technique is also applicable to other setups where the mechanical frequency can be varied and hence the wave function of the mechanical oscillator expanded~\cite{Chang2010a, Singh2010b}.

We discuss now the different stages of the protocol using levitated nano-spheres~\cite{Romero-Isart2010b, Chang2010, Romero-Isart2011, Barker2010a} trapped within an optical cavity (Fig.~\ref{Fig:Fig1}a). The optomechanical coupling is given by $g=x_0 \sqrt{n_\text{ph}} \epsilon_c k_c^2 c V /(4 V_c)$, where $\epsilon_c \equiv 3 \text{Re} \left[ (\epsilon_r-1)/(\epsilon_r+2) \right]$ depends on the relative dielectric constant $\epsilon_r$, $c$ is the vacuum speed of light, $V$ the volume for a sphere of radius $R$ and mass $m$, $V_c$ the cavity volume, and $n_\text{ph}$ the cavity photon number in the steady state. The decoherence rate of the center of mass motion is dominated by light scattering and is given by $\Gamma=\Lambda_\text{sc} x_0^2$, with a localization rate $\Lambda_\text{sc}=\epsilon^2_c n_\text{ph} c V^2 k_c^6 /(6 \pi V_c)$~\cite{Romero-Isart2011,Chang2010}. The decay rate of the cavity also has a contribution due to light scattering given by $\kappa_\text{sc} = \epsilon_c^2 V^2 k_c^4 c /(16 \pi V_c)$. Sideband cooling of the center of mass motion allows the preparation of thermal states to a final number occupation given by $\bar n \approx [\kappa/(4 \omega_t)]^2+ \mathcal{C}_l^{-1}$~\cite{Cooling}, where backaction heating contributes with $\mathcal{C}_l \sim c/(k_c^2 V_c \kappa)$ for a levitated object at very low pressures. Moderate cooling along the other directions is also applied to keep the trap stable at low pressures and reduce the position fluctuations during the time of flight.  After cooling, the harmonic trap is switched off, the object falls (see Fig.~\ref{Fig:Fig1}b), and the state evolves freely according to
\be \label{eq:MEGamma}
\dot{\hat{\rho}} = \frac{\im}{2 m \hbar } \comm{\hat \rho}{\pop^2} - \Lambda \comm{\xop}{\comm{\xop}{\hat \rho}}.
\ee
The position-localization dissipation part of this master equation describes standard decoherence processes such as scattering of air molecules and emission, absorption, and scattering of black body radiation~\cite{BooksDec} with the total localization rate $\Lambda_\text{sd}$ given below. We remark that decoherence due to light scattering is absent during the time of flight since the lasers are switched off. Since both the initial state and the master equation Eq.~\eqref{eq:MEGamma} are Gaussian, the evolved density matrix can be fully determined by computing the moments
$\avg{\xop^2(t)}$,  $\avg{\pop^2(t)}$, and
$\avg{\xop(t)\pop(t)}$, 
where $\avg{\xop^2(0)} =(2 \bar n+1) x_0^2 $, and $\avg{\pop^2(0)} = (2 \bar n+1)\hbar^2/(4x_0^2)$.  The spatial coherence length $\xi_l$, obtained by noticing that $\bra{-x/2}\hat \rho \ket{x/2} \propto \exp \left[ - x^2/\xi_l^2\right]$, is given by $\xi^2_l = 8 \avg{\xop^2} \avg{\mathcal{P}}^2$, where $\avg{\mathcal{P}}$ is the mean value of the parity operator.

After an expansion of duration $t_1$, a second cavity is used to implement an optomechanical double slit (Fig.~\ref{Fig:Fig1}c). To this end, the setup is aligned such that the object passes through a small high-finesse optical cavity at an antinode of the cavity mode. Simultaneously, a pulse of length $\tau \approx 2 \pi/\kappa$ is fed into the cavity such that a short interaction is triggered. Note that during this interaction, standard decoherence and, in particular, light scattering decoheres the state of the system with a rate given by $\bar \Gamma = \Lambda_\text{sc} \avg{\xop^2(t_1)}$. This can be taken into account by adding the corresponding contribution of time $\tau$ to the moments of the Gaussian state before the measurement.  

Linear pulsed optomechanics has been discussed recently for tomography and cooling applications~\cite{Vanner2010a}. Here, we extend these results to the case of the quadratic coupling (see also~\cite{Jacobs2009}). The interaction Hamiltonian, in the displaced frame and in the rotating frame with the resonant laser frequency, is given by $
\hat H= \pop^2/(2 m) + \hbar \bar g_{q} \sqrt{n_\text{ph}} \tilde{x}^2 + \hbar \bar g_{q} (\adop+\aop) \tilde x^2$. A key remark is that, at this stage, the dimensionless position operator is defined as $\tilde x= \xop/\sigma(t_1)$ (hereafter we define $\sigma^2 \equiv \sigma^2(t_1)= x_0^2+ \hbar^2 t_1^2/(4x_0^2m^2)$). Then $\bar g_q \equiv g_q (\sigma/x_0)^2$ is the quadratic optomechanical coupling enhanced by the enlarged wavefunction. Note that, as mentioned above, the kinetic term can be neglected since $\tau \avg{\pop^2}/(2 m \hbar) \approx \omega_t \tau/4\ll 1$ for short cavities where $\kappa \gg \omega_t$. The squared position measurement is performed by measuring the integrated output quadrature of the light field $\pop_L \equiv\int_0^\tau dt [ \adop_\text{out} (t) + \aop_\text{out} (t) ]/\sqrt{\tau}$. Using the input-output formalism, $\aop_\text{out}(t) + \aop_\text{in}(t) = \sqrt{2 \kappa} (\aop + \sqrt{n_\text{ph}})$, one obtains that $\avg{\pop_L}=\chi \avg{\tilde x^2}$ and $\avg{\pop^2_L-\avg{\pop_L}^2}=1/2+\chi^2 \avg{\tilde x^4-\avg{\tilde x^2}^2}$ (we assume a coherent drive such that the optical input phase noise is $1/2$). Therefore, the measurement strength of the squared position measurement is defined as $\chi = 2 \sqrt{ \bar{\mathcal{C}}_q}$ (the physical parameters are chosen such that $\tau \approx 1/\bar \Gamma \approx 2 \pi/\kappa$, see below). Note that the measurement strength is intimately related to the enhanced non-linear cooperativity, see Eq.~\eqref{eq:cooperativity}. The generalized measurement operator for the measurement outcome $p_L$ of the integrated optical phase $\pop_L$ is given by
\be \label{eq:measurement}
\hat{\mathcal{M}} = \exp{ \left[ - \im \phi \tilde x^2 -\left( p_L - \chi \tilde x^2 \right)^2\right]},
\ee
where $\phi= \bar g_{q} \sqrt{n_\text{ph}} \tau $ is the phase accumulated during the interaction with the classical field. The density matrix after the measurement is described by $\hat \rho(t_2+\tau)=\hat{\mathcal{M}} \hat \rho \hat{\mathcal{M}}^\dagger/ \tr[ \hat{\mathcal{M}} \hat \rho \hat{\mathcal{M}}^\dagger] $. 
The action of the measurement operator, Eq.~\eqref{eq:measurement}, is to prepare a superposition of two wave  packets separated by a distance $d=2 \sigma \sqrt{p_L/\chi}$, and a width given by approximately $\sigma_2 \sim \sigma /(4\sqrt{p_L \chi}) = \sigma^2/(2 d \chi)$. This can be intuitively understood as a consequence of the projective nature of the pulsed measurement~\cite{Vanner2010a}: for the ideal case, this measurement prepares the system in an eigenstate of the $\xop^2$ operator, which for a pure initial state with even parity is of the type $|x\rangle+|-x\rangle$, \ie~a coherent spatial superposition. The separation of the wave packets, $d$, determined by the outcome of the measurement, represents the effective slit separation. In order to prepare (with high probability) and resolve the peaks of the superposition state, one requires $\sigma > d > 2 \sigma_2$. This sets up an upper bound $d^\text{a}_\text{max}\equiv \sigma$ and a lower bound $d_\text{min}\equiv \sigma \sqrt{2/\chi}$ for $d$. A second upper bound is provided by the decoherence during the expansion of the wave function; we demand $d < d^\text{b}_\text{max}\equiv\xi_l$.  Finally, the total number of photons $n_\text{ph}$ used in the pulse and the time of flight $t_1$ are fixed by enforcing that $\phi$ compensates the complex phase accumulated during the time of flight, which is given by  $\sim \avg{\xop(t_1) \pop(t_1) + \pop(t_1) \xop(t_1)}/(4 \hbar)$, as well as by fulfilling the condition  $\tau \approx 1/\bar \Gamma \approx 2 \pi /\kappa$.  This corresponds to choosing $n_\text{ph} \approx (2 \bar n+1)/[32 \pi \mathcal{C}_l (k_c x_0)^2]$ and $t^2_1 \approx 16 \kappa \mathcal{C}_l k_c^2/[\omega^2 (2 \bar n+1)^2 \Lambda_\text{sc}/n_\text{ph}]$. 

\begin{figure}
\includegraphics[width=\linewidth]{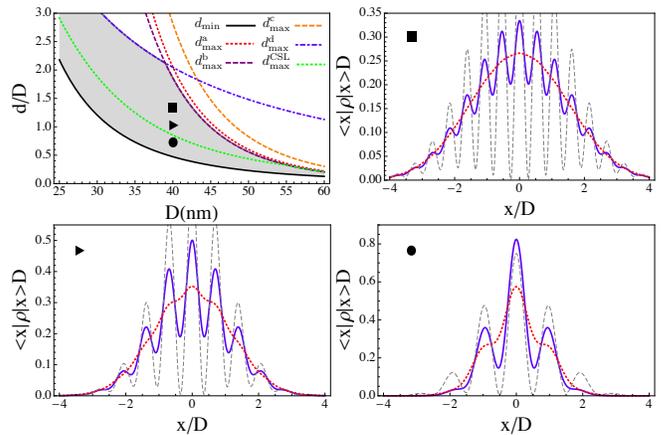}
\caption{(Color online) The operational parameter regime for the optomechanical double slit distance $d$ and the the diameter of the sphere $D$ is plotted (see legend for the lower and upper bounds). The simulation of the interference pattern is computed for a sphere of $40$ nm and $d=0.7 D$ (circle), $d=D$ (triangle), and $d=1.3 D$ (square) in units of $D$. The solid blue (dashed grey) line is the simulated interference pattern with (without) standard decoherence. The dotted red line is the interference pattern in the presence of the CSL model with $\lambda=10^4 \lambda_0$ $\text{s}^{-1}$ (the upper bound $d^\text{CSL}_\text{max}$ in the operational parameter plot provided by the CSL model is also shown, see legend).  Experimental parameters for the environmental conditions: $P=10^{-16}$~Torr, 
$T_e=4.5$~K, 
$\text{Im} \left [(\epsilon_\text{bb}-1)(\epsilon_\text{bb}+2)\right]=0.1$,
 $\text{Re}[\epsilon_\text{bb}]=2.3$, 
 $\bar n=0.1$; for the cavity: finesse $\mathcal{F}=1.3 \times 10^5$, 
 length $2$~$\mu$m,  waist $=1.5$ $\mu$m, $\lambda_c=1064$ nm; and for a silica sphere:
 $\epsilon_r=2.1 + \im 2.5 \times10^{-10}$, density $=2201$ Kg/$\text{m}^3$,
 $\omega_t/2\pi=135$ KHz, and
 $\delta x=10$ nm. Using this, for a sphere of $40$ nm and slit length $d=D$, one obtains 
 $\kappa/2 \pi=\tau^{-1}= 2.8 \times 10^8 $ Hz,  $\mathcal{C}_l=1500$, $n_\text{ph}=272$,  $T_i=206$ K, $t_1=3.3$ ms, $t_2=125$ ms, and $\sigma/x_0=2928$.}
\label{Fig:Fig2}
\end{figure}

After the preparation of the superposition state by the pulsed interaction, the particle falls freely during another time of flight of duration $t_2$. An interference pattern in the mean value of the position is formed with fringes separated by a distance   $x_f= 2 \pi \hbar t_{2}/(m d)$. The final step of the protocol is thus to perform a position measurement of the center of mass (Fig.~\ref{Fig:Fig2}d). This requires a resolution $\delta x < x_f$, providing a third upper bound for $d$,  $d_\text{max}^\text{c}\equiv2 \pi \hbar t_2/(m \delta x)$. Note that sufficiently long time $t_2 \sim m \sigma^2/(\hbar \chi)$ is needed to guarantee the overlap of the two wave packets. The effect of standard decoherence on the visibility of the interference pattern can be obtained by solving the evolution of the position distribution for a non-Gaussian state under the evolution of~Eq.\eqref{eq:MEGamma}. This is given by the closed expression~\cite{Ghirardi1986}
\be \label{eq:blurring}
\bra{x}\hat  \rho(t) \ket{x} =\int_{-\infty}^\infty dy  \frac{e^{-y^2/\sigma_b^2(t)}}{\sigma_b(t) \sqrt{\pi}}\bra{x+y} \hat \rho_{\Lambda=0}(t)\ket{x+y},
\ee
where $\hat \rho_{\Lambda=0}(t)$ is the state obtained with the evolution due to the Schr\"odinger equation only, that is with $\Lambda=0$. As observed in \eqref{eq:blurring}, the effect of decoherence is to blur the position distribution with a blurring coefficient given by
$\sigma_b(t)= 2\hbar m^{-1} \sqrt{t_2^3 \Lambda/3}$. Therefore, the fringes separated by a distance $x_f$ will be visible provided $x_f > \sigma_b(t_2)/2$, which provides the fourth upper bound $d_\text{max}^\text{d}\equiv\pi\sqrt{3  /(t_2 \Lambda_\text{sd})}/2$.
Putting everything together, the operational regime for the experiment modelled here is given by
$ d_\text{min} <d< \min \left\{ d^\text{a}_\text{max},d^\text{b}_\text{max},d^\text{c}_\text{max},d^\text{d}_\text{max} \right\}$.

We now address the experimental conditions required for this experiment. 
The localization rate for black body radiation $\Lambda_\text{bb}$ has contributions due to scattering $\Lambda_\text{bb,sc} \propto R^6 T_{e}^{9}  \text{Re} \left [(\epsilon_\text{bb}-1)/(\epsilon_\text{bb}+2) \right]^2$, and  emission(absorption) of blackbody radiation $\Lambda_\text{bb,e(a)} \propto  R^3  T_{i(e)}^6 \text{Im} \left [(\epsilon_\text{bb}-1)/(\epsilon_\text{bb}+2) \right]$, see~\cite{BooksDec, Chang2010} for the exact expressions.  $\epsilon_\text{bb}$ is the average relative permittivity, which is assumed to be constant across the relevant blackbody spectrum, and $T_{i(e)}$ is the internal (environmental) temperature. $T_{i}$ at very low pressure can be computed using the balance between the emitted blackbody power and the light absorption during the optical cooling and trapping~\cite{Chang2010}.  
Second, decoherence due to air molecules is described by the master equation Eq.~\eqref{eq:MEGamma}~\footnote{This is valid for separations smaller than the de Broglie wavelength of the air molecules. For larger separations, this master equation is also a good approximation since one can first relate it to a GRW type master equation, see ~\cite{Gallis1990}, and then use the infinite frequency limit~\cite{Ghirardi1986}.}, with the parameter given by~\cite{BooksDec} $\Lambda_\text{air}= 8 \sqrt{2 \pi} m_a \bar v  P R^2/(3 \sqrt{3}\hbar^2)$, where $P$ is the air pressure, $m_a$ is the mass of the air molecules and $\bar v$ their thermal velocity. The total standard decoherence rate is thus given by $\Lambda_\text{sd}= \Lambda_\text{bb}+ \Lambda_\text{air}$. The overall performance of this challenging experiment is mainly limited by the quality of the cavity used in the measurement and the vacuum and temperature conditions required for the the environment. In particular, very good vacuum conditions are needed to keep the coherence of these fragile states. Note however that pressures down to $10^{-17}$ Torr at cryogenic temperatures of $T=4.5$~K were reported in~\cite{Gabrielse1990}. Extremely good cavities are needed in order to obtain a large cooperativity $\mathcal{C}_l$, for instance, consider fiber-based Fabry-Perot cavities of length of $2$ $\mu$m and finesse $\mathcal{F}\approx 1.3 \times 10^5$ as discussed in~\cite{Hunger2010}. In Fig.~\ref{Fig:Fig2} the operational parameter regime is shown for different sphere sizes and superposition distances with the particular set of experimental parameters given in the caption. The interference pattern simulated by solving the master equation numerically, which describes the evolution of the state during the experiment, is also plotted. Spheres of $\sim 40$ nm with a mass of $\sim10^7$ amu can be prepared in a superposition of locations separated by a distance equal to their diameter. In principle, the scheme can be applied to even larger objects albeit with further constraints on the experimental parameters.

To conclude, we shall discuss the application of using this experiment to test theories beyond quantum mechanics that provide an objective collapse of the wavefunction for sufficiently large objects. In particular, we focus on the paradigmatic model associated to Ghirardi-Rimini-Weber-Pearle, see \cite{Bassi2003} and references therein, denoted as the continuous spontaneous localization model (CSL). This  theory is derived  by adding a non-linear stochastic term to the Schr\"odinger equation. The model recovers all the phenomenology of quantum mechanics for elementary particles but predicts a fast localization (collapse) of the wavefunction for larger objects. This comes at the price of introducing two phenomenological constants given by  $\alpha^{-1/2} \approx 10^{-7}$ m (related to the localization extension) and $\lambda_0 \approx 2.2 \times 10^{-17}$ $\text{s}^{-1}$ (related to the intensity of the localization).  For a spherical body~\cite{Collett2003}, the CSL model can be cast into a master equation of the form of Eq.~\eqref{eq:MEGamma} with 
$\Lambda_\text{CSL} =  m^2 \lambda_0 \alpha f(\sqrt{\alpha} R)/(2m^2_0)$,
where $m_0$ is the mass of a nucleon, and the function $f(x)$ defined in \cite{Collett2003} has the following limits:
$f(1)\approx 0.62$, $f(x \ll 1) =1$, and $f(x \gg 1)\approx 6 x^{-4}$. Recently, Adler~\cite{Adler2007} has reexamined the CSL theory and, by considering the collapse of the wave function at the latent image formation level, he predicted a significantly larger value for $\lambda_0$, namely  $\lambda_\text{A}=2 \times 10^{9 \pm 2} \lambda_0$. This prediction cannot be tested by current experiments~\cite{Adler2009}. In Fig.~\ref{Fig:Fig2} we show however that a possible CSL process would have a strong impact on our experiment already for $\lambda =10^{4} \lambda_0 $ (see the upper bound $d^\text{CSL}_\text{max}$ provided by the blurring of the interference pattern). The effect is also clearly visible in the simulation of the interference pattern. Thus, the experiment proposed here puts unprecedented bounds for one of the most studied collapse models and even challenges the recent theoretical prediction given by Adler~\footnote{During the submission of this article, we became aware of a theoretical analysis to test spontaneous localization models with matter-wave interferometry~\cite{Nimmrichter2011a}.}. Finally we note that our scheme allows to prepare superpositions of macroscopically distinct spatial states of a massive object. In combination with the specific time-of-flight evolution this may provide a rigorous experimental test of some of the crucial assumptions of macrorealism~\cite{Leggett1985}.

We are grateful to M. D. Lukin, E. M. Kessler, and F. Pastawski for stimulating discussions. We acknowledge support of Alexander von Humboldt Stiftung, ENB (Project QCCC), Caixa Manresa, EU (AQUTE, MINOS, Q-ESSENCE, Marie Curie), FWF (START, FOQUS), ERC (StG QOM), and FQXi.

\bibliographystyle{apsrev}

\end{document}